\documentclass[aps,prl,reprint,superscriptaddress,nofootinbib]{revtex4-2}
\usepackage{amsmath}
\usepackage{mathtools}
\usepackage{amsfonts,amssymb}
\usepackage[dvipsnames]{xcolor}
\usepackage{tikz}
\usepackage{graphicx}
\usepackage{hyperref}
\usepackage{braket}
\usepackage{calc}
\hypersetup{
 bookmarksnumbered=true,
 colorlinks=true,
 linkcolor=[rgb]{0.098,0.098,0.439},
 citecolor=[rgb]{0.098,0.098,0.439},
 urlcolor=[rgb]{0.098,0.098,0.439}
  }

\begin{document}

\title{Synaptic Field Theory for Neural Networks}

\author{Donghee Lee}
\email{dhlee641@kaist.ac.kr}
\author{Hye-Sung Lee}
\email{hyesung.lee@kaist.ac.kr}
\author{Jaeok Yi}
\email{wodhr1541@kaist.ac.kr}
\affiliation{Department of Physics, Korea Advanced Institute of Science and Technology, Daejeon 34141, Korea}

\date{March 2025}

\begin{abstract}
Theoretical understanding of deep learning remains elusive despite its empirical success. In this study, we propose a novel “synaptic field theory” that describes the training dynamics of synaptic weights and biases in the continuum limit. Unlike previous approaches, our framework treats synaptic weights and biases as fields and interprets their indices as spatial coordinates, with the training data acting as external sources. This perspective offers new insights into the fundamental mechanisms of deep learning and suggests a pathway for leveraging well-established field-theoretic techniques to study neural network training.
\end{abstract}

\maketitle

\textit{Introduction---} The remarkable success of deep learning in a wide range of applications has triggered intense efforts to understand its core mechanisms. While deep neural networks are highly effective at extracting rich features, the fundamental reasons for their success remain elusive. Numerous strategies have been proposed to clarify these mysteries; for general reviews, see Refs.~\cite{LeCun:2015dl, fan2021interpretability, li2022interpretable, bahri2020statistical}. One of these strategies is to utilize field theory, which serves as a fundamental framework in physics. 

Field-theoretic approaches have been proposed in numerous studies \cite{Krippendorf:2022hzj, hashimoto2018deep, Hashimoto:2018bnb, Hashimoto:2019bih, gan2017holography, lee2017deep, antognini2019finite, hanin2019finite, huang2019dynamics, dyer2019asymptotics, yaida2019non, naveh2021predicting, seroussi2021separation, aitken2020asymptotics, andreassen2020asymptotics, halverson2021neural, zavatone2021asymptotics, naveh2021self, hanin2022correlation, yaida2022meta, demirtas2024neural, Erbin:2021kqf, Erbin:2022lls, ringel2025statistical, Buice_2013, sonoda2018transportanalysisinfinitelydeep, Helias2020statistical, halverson2021building, maiti2023symmetry, Halverson:2024axc, Grosvenor:2021eol, Gukov:2024buj, PhysRevResearch.3.023034, howard2024bayesian, Aarts:2024rsl, PhysRevE.75.051919, Berman:2022uov, Berman:2023rqb,   lee2020quantum, Jia2019quantum, carleo2017solving, PhysRevX.7.021021,   Bachtis:2021xoh,    2023PhyA.61228492Z, 2023PTEP.2023f3A01Z, Banta:2023kqe, Vanchurin:2024zpn, Segadlo_2022, bondesan2021hintonsneuralnetworkquantum}. For instance, in Refs.~\cite{hashimoto2018deep, Hashimoto:2018bnb, Hashimoto:2019bih, gan2017holography}, the authors propose that deep neural networks reflect the structure of the AdS/CFT correspondence by using neural networks to learn and reconstruct the AdS metric from data. The neural network/Gaussian process correspondence \cite{neal1996priors, williams1996computing, matthews2018gaussian, novak2018bayesian, garriga2018deep, yang2019scaling, yang2019wide, yang2020tensor, jacot2020neuraltangentkernelconvergence}  has prompted numerous field-theoretic studies, and finite-width effects have been extensively investigated using various approaches \cite{lee2017deep, antognini2019finite, hanin2019finite, huang2019dynamics, dyer2019asymptotics, yaida2019non, naveh2021predicting, seroussi2021separation, aitken2020asymptotics, andreassen2020asymptotics, halverson2021neural, zavatone2021asymptotics, naveh2021self, hanin2022correlation, yaida2022meta, demirtas2024neural, Erbin:2021kqf, Erbin:2022lls, ringel2025statistical}. There are other intriguing works, too \cite{sonoda2018transportanalysisinfinitelydeep, Helias2020statistical, halverson2021building, maiti2023symmetry, Halverson:2024axc, Grosvenor:2021eol,  lee2020quantum, Jia2019quantum, carleo2017solving, PhysRevX.7.021021, PhysRevE.75.051919,   Bachtis:2021xoh, Buice_2013,   PhysRevResearch.3.023034, 2023PhyA.61228492Z, 2023PTEP.2023f3A01Z, Banta:2023kqe, Gukov:2024buj, howard2024bayesian, Aarts:2024rsl, Berman:2022uov, Berman:2023rqb, Vanchurin:2024zpn, Segadlo_2022, bondesan2021hintonsneuralnetworkquantum}. Among these studies, some have emphasized the role of symmetry \cite{halverson2021building, maiti2023symmetry, ringel2025statistical, Halverson:2024axc, Helias2020statistical}, while others have applied statistical physics \cite{PhysRevE.75.051919, Erbin:2021kqf, ringel2025statistical, howard2024bayesian, Buice_2013, Erbin:2022lls, Grosvenor:2021eol, PhysRevResearch.3.023034, Berman:2022uov, Berman:2023rqb, howard2024bayesian, Gukov:2024buj, Aarts:2024rsl}. Especially, the connection to renormalization group transformations has been explored in several works \cite{PhysRevE.75.051919, Erbin:2021kqf, Erbin:2022lls, Grosvenor:2021eol, PhysRevResearch.3.023034, Berman:2022uov, Berman:2023rqb, howard2024bayesian, Gukov:2024buj, Aarts:2024rsl}.

Krippendorf and Spannowsky (KS) made an important observation, identifying a duality connecting neural network and cosmological dynamics \cite{Krippendorf:2022hzj}.
To substantiate this duality, they took an effective field theory (EFT) approach, and demonstrated that the time evolution of the network’s outputs can be mapped onto that of a cosmological system in the limit where the neural tangent kernel (NTK) becomes constant \cite{jacot2020neuraltangentkernelconvergence}.

In this regime, they related cosmological parameters, such as the Hubble parameter, to training parameters of the network, and pointed out a duality between neural networks and de Sitter (dS) space.\footnote{Although KS used the term ``vacuum energy dominated universe,'' we use dS space for simplicity. While a universe dominated by vacuum energy can indeed be described as dS space, not all dS spacetimes correspond to such a universe.}

Nevertheless, we must develop a more fundamental theory that treats the parameters of neural networks—synaptic weights and biases— directly for various reasons. Here, we introduce a field‑theoretic framework that elevates them—the network's fundamental degrees of freedom—to dynamical fields. A theory formulated in terms of these neural network parameters can be viewed as a UV description of the training dynamics.

Because the parameters are numerous, their collective behavior is naturally captured by a continuum description. We therefore construct a ``synaptic field theory,'' the first formalism to treat synapses explicitly as fields. Although taking the continuum limit introduces technical challenges, the conceptual bridge between neural networks and field theory remains clear and robust. This framework brings the full machinery of field theory to neural network analysis and promises new insights into network dynamics.

\vspace{2mm}
\textit{Neural Network Theory and Previous Works---}
A deep neural network (DNN) comprises layers of neurons, each connected by weights \(W_{ij}^{(m)}\) and biases \(b_i^{(m)}\equiv W_{i0}^{(m)}\). For \((m+1)\)-th layer, the neuron \(h_i^{(m+1)}\) depends on all neurons \(h_j^{(m)}\) in the previous layer according to  
\begin{equation} \label{eq: recursive relation}
h_i^{(m+1)} 
= \sigma\Big(\sum_j W_{ij}^{(m)}\,h_j^{(m)}\Big),
\end{equation}
where \(\sigma\) is a non-polynomial activation function \cite{lederer2021activationfunctionsartificialneural}. Here, the sum over $j$ runs from 0 to $N$, and $h_0^{(m)} = 1$ for all $m$ which runs from 0 to $M$. Iterating Eq.~\eqref{eq: recursive relation} from the input (initial layer) \(X=h^{(0)}\) to the output (final layer) \(Z=h^{(M)}\) defines the forward pass of the network.

Training a DNN involves adjusting \(W_{ij}^{(m)}\) to minimize the quadratic cost function 
\begin{equation}
      C=\sum_{i,l}\Big(Y_i^{[l]} - Z_i^{[l]}\Big)^2, \label{eq:costfunction}
\end{equation}
where \((X_i^{[l]}, Y_i^{[l]})\) are the training inputs and desired outputs, and \(Z_i^{[l]}\) is the DNN prediction for \(X_i^{[l]}\). The index $l$ labels individual training examples. Gradient descent updates weights and biases according to
\begin{equation} \label{eq:gradient_descent} \Delta W_{ij,\ T}^{(m)} = -\eta\frac{\partial C}{\partial W_{ij}^{(m)}}, \end{equation}
where $\Delta W_T = W_{T+1} - W_T$, with $W_T$ representing the value of $W$ at training step $T$, and $\eta$ denoting the step size. The derivatives on the right-hand side of the equation are evaluated at the weights and biases from the previous training step $T$.
Interpreting the discrete step \(T\) as a continuous time \(t\) yields the differential equation  
\begin{equation}
    \dot{W} 
    = -\eta\frac{\partial C}{\partial W},
    \label{eq:timederivativegd}
\end{equation}
with \(W\) representing all \(W_{ij}\).

Although sometimes referred to as an equation of motion for gradient descent, Eq.~\eqref{eq:timederivativegd}, cannot be derived from the least action principle. Nonetheless, it can be viewed as the high-viscosity limit (\(\gamma \dot W \gg \ddot W\)) of the second-order differential equation\footnote{This equation can also be interpreted as describing a neural network with momentum~\cite{Krippendorf:2022hzj,ruder2017overviewgradientdescentoptimization, Lee2020Wide}.}~\cite{Parisi1988statistical, RevModPhys.65.499}
\begin{equation}
    \ddot{W} 
    + \gamma\dot{W} 
    + \frac{\partial C}{\partial W} 
    = 0.
    \label{eq:neweq}
\end{equation}

This equation resembles the equation of motion for a scalar field $\phi$ in curved spacetime,
\begin{equation}
\ddot{\phi} + 3H \dot{\phi} + \frac{\partial V}{\partial \phi} = 0,
\end{equation}
when we identify $W$ with $\phi$, $\gamma$ with $3H$, and $C$ with $V$.

Because $\gamma$ is constant in typical training, Eq.~\eqref{eq:neweq} suggests a connection between neural network training and dynamics in dS space. Motivated by this analogy, KS proposed that the time evolution of the network output $Z$ obeys
\begin{equation}
\ddot{Z} - \beta \dot{Z} + \Theta \frac{\partial C}{\partial Z} = 0,
\label{eq:KS}
\end{equation}
where $\beta$ is related to $\gamma$ and $\Theta$ denotes the NTK. Because $\Theta$ becomes constant in a certain limit, KS argued that Eq.~\eqref{eq:KS} establishes a duality between the time evolution of $Z$ and cosmological dynamics~\cite{Krippendorf:2022hzj}.

To discuss further, KS exploited the EFT approach\footnote{Likewise, several studies have sought to develop EFT frameworks for deep learning~\cite{Banta:2023kqe, howard2024bayesian, PhysRevResearch.3.023034, Krippendorf:2022hzj}.} and conducted a simulation analysis. EFT is indeed a highly promising methodology, but the fundamental degrees of freedom that actually change during training are the neural network parameters, and developing a theory that directly addresses them is worth pursuing. In particular, the analysis by KS rests on the assumption that the empirical NTK~\cite{samarin2020empiricalneuraltangentkernel, fort2020deeplearningversuskernel, novak2019neuraltangentsfasteasy} accurately approximates network dynamics during training. While the NTK captures how parameter variations influence the output, it becomes constant only in restricted cases. Consequently, a more careful examination of how parameters affect the output is required, and constructing a theoretical framework that focuses on the parameters themselves is especially meaningful.

In the following section, we adopt a more fundamental approach, introducing a field‑theoretic description of neural network parameters that begins with their microscopic training dynamics. This framework will also reveal a connection to cosmological dynamics, akin to the one explored by KS~\cite{Krippendorf:2022hzj}.

\begin{figure*}[t]
\begin{tikzpicture}[scale=0.6, every node/.style={scale=0.6}]
    \node at (-7.5,0) {\includegraphics[width=.65\linewidth]{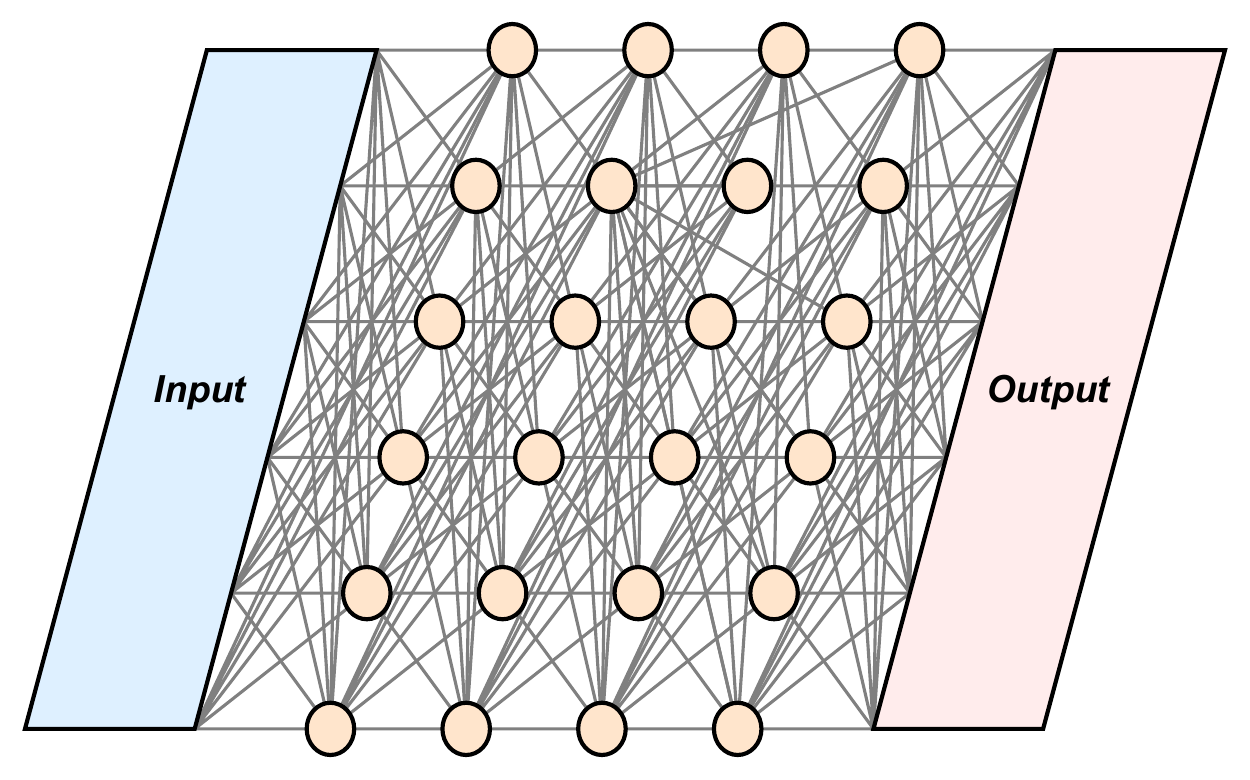}};
    \node at (7.5,0) {\includegraphics[width=.65\linewidth]{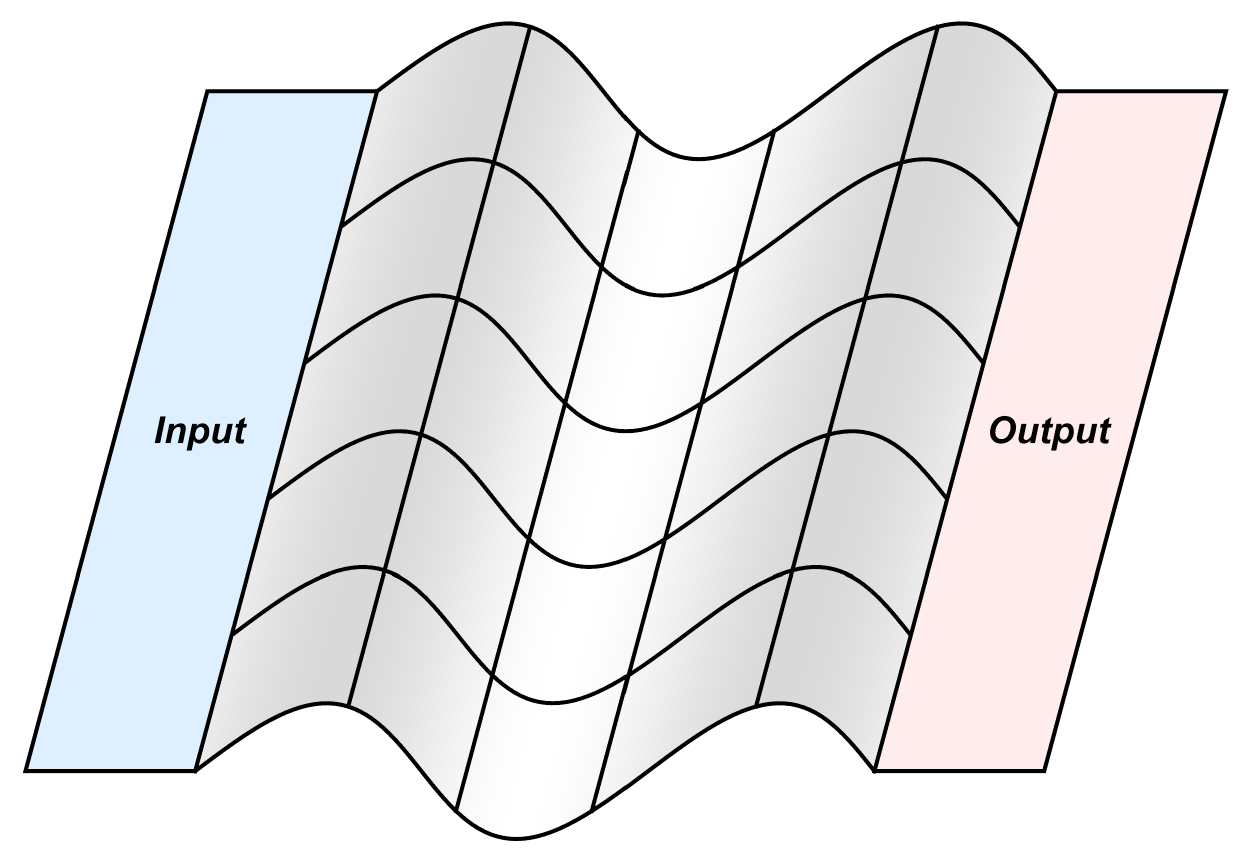}};
    \draw[ultra thick, <->] (-1+1/4,0) -- (1-1/4,0);
    \node at (-7.5,-4.5) {\Large $\displaystyle C=\sum \left(Y_i^{[l]} - Z_i^{[l]}[W_{ij}^{(m)}]\right)^2$ \qquad \textcolor{gray}{\raisebox{3pt}{\rule{1cm}{2pt}}} $\displaystyle W_{ij}^{(m)}$ };
    \node at (7.5,-4.5) {\Large $\displaystyle S=\int d^{d+1}x \ \sqrt{-g} \; \mathcal L[w(t,\mathbf x)] $ \qquad \textcolor{gray}{$\blacksquare$} $ w(t,\mathbf x)$};
\end{tikzpicture}
\caption{Visualization of synaptic field theory. We introduce synaptic field theory by interpreting the collective behavior of synaptic weights and biases as a field $w(t,\mathbf x)$ interacting with external sources driven by training datasets. The space of the synaptic field is defined by indices representing neuron positions and their synaptic connections, providing a structured framework to describe neural network dynamics in a field-theoretic manner.}
\label{fig:general}
\end{figure*}

\vspace{2mm}
\textit{Synaptic Field Theory---} While the original equation of motion for the synaptic weights~\eqref{eq:timederivativegd} is not derived from the least action principle, Eq.~\eqref{eq:neweq} can be obtained from the action
\begin{equation}
    S 
    = \int dt \ e^{\gamma t}\left[\frac12 \dot{W}^2 - C\right] , \label{eq:actiondc}
\end{equation}    
where \(\frac12 \dot{W}^2\) denotes \(\sum_{i,j,m}\tfrac12 \dot{W}_{ij}^{(m)2}\). The terms inside the brackets resemble a matter Lagrangian, as they follow the typical kinetic minus potential structure.    
If a suitable continuum limit of Eq.~\eqref{eq:actiondc} exists, or if the (matter) Lagrangian can be recast as an integral of a Lagrangian density,
\begin{equation}
    L[W(t)]
    = \frac12 \dot{W}^2 - C
    = \int d^d x \,\mathcal{L}[w(t,\mathbf{x})], \label{eq:cont}
\end{equation}
then the action can be written as
\begin{equation}
    S
    = \int d^{d+1}x \ e^{\gamma t} \mathcal{L}[w(t,\mathbf x)].
\end{equation}

Identifying $ e^{\gamma t} =  \sqrt{-g} $ reveals a connection between neural networks and a field theory in curved spacetime. The exponential factor \(\sqrt{-g}\) appears in the universe dominated by the cosmological constant. In $(d+1)$-dimensional dS spacetime, the Hubble parameter \(H\) becomes a positive constant and the metric tensor takes the form $\sqrt{-g} = e^{dHt}$. Matching exponents identifies constant Hubble parameter \(H = \gamma/d\) and indicates the connection to dS space.
    
Notably, the cost function in Eq. \eqref{eq:costfunction} sums over synaptic weight and bias indices. To construct a continuum theory, it is natural to take the continuum limit of these indices, replacing the summation with an integral. We call this theory, defined on the space based on these indices, ``synaptic field theory." Figure~\ref{fig:general} schematically illustrates this.

\begin{table}[b]
    \centering
    \begin{tabular}{c|c}
       \textbf{Neural Network} & \textbf{Synaptic Field Theory} \\ \hline
       Weight $W_{ij}^{(m)}$ & Field $w(t,\mathbf x)$ \\
       Training examples $(X,Y)$ & External sources $J, K, \cdots$ \\
       Indices $i,j,m$ & Space $\mathbf x$ \\ 
       Training step $T$ & Time $t$ \\
       Cost function $C$ & Lagrangian $ L$ \\
       Step size $\eta$ & Hubble parameter $H$ 
    \end{tabular}
    \caption{A dictionary relating neural network components to the synaptic field is presented.}
    \label{tab:my_label}
\end{table}

If the activation function admits an infinite‑series expansion (as the sigmoid does), the cost function can likewise be expressed as an infinite series in the weights:
\begin{multline}
C =  \sum  J_{1\ i_1j_1}^{(m_1)} W_{i_1j_1}^{(m_1)}   \\ + \sum  J_{2\ i_1j_1i_2j_2}^{(m_1m_2)} W_{i_1j_1}^{(m_1)} W_{i_2j_2}^{(m_2)} +\cdots .
\end{multline}
The coefficients $J_{1\ i_1j_1}^{(m_1)}$ and $J_{2\ i_1j_1i_2j_2}^{(m_1m_2)}$ depend on the data set, and the expression is defined up to an additive constant. Repeated indices are summed.

Taking the continuum limit—replacing discrete indices by three‑dimensional spatial coordinates—yields
\begin{multline}
    L \supset   \int  d^3 \mathbf x \; J_1(\mathbf x) w(t,\mathbf x)  \\  + \int  d^3 \mathbf x d^3 \mathbf y \; J_2(\mathbf x, \mathbf y) w(t,\mathbf x) w(t,\mathbf y) + \cdots .
\label{eq:clgen}
\end{multline}
Here, $J_1$ and $J_2$ act as external sources determined by the training examples.
Table~\ref{tab:my_label} summarizes how the neural network components map onto elements of the synaptic field theory.

\begin{figure*}[t]
\begin{tikzpicture}[scale=0.65, every node/.style={scale=0.65}]
    
    \node[circle, draw, minimum size=1.3cm, thick,fill=RoyalBlue!6] (X1) at (0-1, 5) {$X_{2N-1}$};
    \node at (0-1,3.6) {\huge $\displaystyle \vdots$};
    \node[circle, draw, minimum size=1.3cm,thick,fill=RoyalBlue!6] (X3) at (0-1, 2) {$X_{3}$};
    \node[circle, draw, minimum size=1.3cm,thick,fill=RoyalBlue!6] (X4) at (0-1, 0) {$X_{1}$};
    \node[circle, draw, minimum size=1.3cm,thick,fill=pink!18] (Z1) at (4-1, 6) {$Z_{2N}$};
     \node at (4-1,4.6) {\huge $\displaystyle \vdots$};
    \node[anchor = west] at (4.7-1,6) {$= \sigma ( W_1 X_1+W_{2N} X_{2N-1})$};
    \node[circle, draw, minimum size=1.3cm,thick,fill=pink!18] (Z3) at (4-1, 3) {$Z_4$};
    \node[anchor = west] at (4.7-1,3) {$= \sigma (W_5 X_5 + W_4 X_3)$};
    \node[circle, draw, minimum size=1.3cm,thick,fill=pink!18] (Z4) at (4-1, 1) {$Z_2$};
    \node[anchor = west] at (4.7-1,1) {$= \sigma (W_3 X_3 + W_2 X_1)$};
    \node[circle, draw, minimum size=1.3cm,thick,fill=pink!18] (Z5) at (4-1, -1) {$Z_{0}$};
    \node[anchor = west] at (4.7-1,-1) {$= Z_{2N}$};
    \draw[->, thick] (X1) -- (Z1) node[midway, fill=white] {\small $W_{2N}$};
    \draw[->, thick] (X1) -- (Z3) node[midway, fill=white] {\small };
    \node[rectangle, draw=white, fill=white, minimum size= 1cm ] at (1,4) {\Large $ \cdots$};
    \draw[->, thick] (X3) -- (Z3) node[midway, fill=white] {\small $W_{4}$};
    \draw[->, thick] (X3) -- (Z4) node[midway, fill=white] {\small $W_{3}$};
    \draw[->, thick] (X4) -- (Z4) node[midway, fill=white] {\small $W_{2}$};
    \draw[->, thick] (X4) -- (Z5) node[midway, fill=white] {\small $W_{1}$};

    \draw[<->, ultra thick] (8.5,3) -- (10,3);
    \node[anchor=west,rectangle,inner sep=3mm] at (11,3) {\Large $ \begin{matrix}  \displaystyle    L  =  \int dx \ \sqrt{-g} \; \bigg [\frac12  [\partial_t w(t,x)]^2  &     \\  & \displaystyle \hspace{-3cm}- \frac12 K(x) [\partial_x w(t,x)]^2   \\ & \displaystyle \hspace{-1cm}- \frac12 J(x) w(t,x)^2   \bigg]  \end{matrix} $
};
         \node at (-3.5,3) {};
         \node at (22,3) {};
\end{tikzpicture}
\caption{Neural network architecture (left) and its continuum limit synaptic field theory (right). This illustrates the correspondence between training examples and external sources, as well as the mapping of discrete indices to spatial coordinates in synaptic field theory.}
\label{fig:example 1}
\end{figure*}

In this approach, it is difficult to endow the synaptic field theory with desirable properties such as locality. The terms in Eq.~\eqref{eq:clgen} are inherently nonlocal, and conventional field theory rarely treats genuinely nonlocal s. Consequently, this nonlocality complicates efforts to formulate a field‑theoretic description of neural networks in dS space.

To find the local Lagrangian, it is important to note that the nonlocality is related to the architecture and indexing convention of the parameters. Equation~\eqref{eq:clgen} is obtained by naively extending a neural network with a typical indexing scheme. Since these indices are mapped to spatial coordinates in the continuum limit, the architecture and indexing convention of the neural network determine the spatial geometry of the synaptic field theory. It is unclear whether typical architectures and indexing conventions yield geometries that are useful for analysis.

In other words, by designing neural networks with specific architectures and assigning appropriate indices, one may obtain a spatial geometry that is easier to analyze or even admits a local action. In the following discussion, we examine two examples in which local synaptic field theories are derived by adopting an appropriate indexing convention for neural networks with a very simple architecture.

\vspace{2mm}
\textit{Examples---} As a first example, consider the perceptron illustrated in Fig.~\ref{fig:example 1}, which has an $N$-component input vector and an $N$-component output vector.
The input and output layers are connected by weights such that only adjacent components interact, and no bias terms are included.
We impose periodic boundary conditions along the width, identifying the $(N+j)$-th neuron with the $j$-th neuron and the $(2N+k)$-th synapse with the $k$-th synapse.
The activation function is linear, $\sigma(x)=px+q$.

The cost function of this perceptron is  
\begin{equation}\label{eq:cost_toy}
    C=\sum_{i,l} \Big[Y_{2i}^{[l]}-\sigma \big(W_{2i}
    ^{}X_{2i-1}^{[l]}+W_{2i+1}^{}X_{2i+1}^{[l]} \big) \Big]^2.
\end{equation}
For notational simplicity, odd-numbered indices label inputs, and even-numbered indices label outputs.\footnote{Typically, weights carry two indices, but restricted connections allow a simplified notation.}  
Because \(C\) is a sum of quadratic polynomials in \(W_{2i}\) and \(W_{2i+1}\), we can expand and shift the weights to recast \(C\), up to constant, as  
\begin{equation}\label{eq: discrete Hamiltonian}
    C=\sum_{i}\bigg[\frac12 K_i(W_{i+1}-W_{i})^2+\frac12 J_i(W_{i})^2\bigg],
\end{equation}
where \(J_i\) and \(K_i\), are coefficients determined by the training data.

This cost function admits the continuum limit under the following heuristic\footnote{To clarify what is meant by the continuum limit, a proper length scale—the lattice spacing—must be introduced. In this correspondence, every occurrence of the lattice spacing in the Lagrangian is assumed to be absorbed into the external source.} correspondences: \begin{equation} \label{eq:continuum}
\begin{split}
\sum_i &\to \int d x, \\
J_i, K_i, W_i, \Delta W_i &\to J( x), K( x), w( t, x),  \partial_ x w(t, x),
\end{split}
\end{equation}
where $\Delta W_i$ denotes $W_{i+1}-W_i$ and \(w\), \(K\), and \(J\) are the continuum analogs of \(W_{i}\), \(K_i\), and \(J_i\), respectively. The continuum limit suggests that the Lagrangian contains the terms
\begin{equation}
    \int dx \ \bigg[\frac12 K(x) [\partial_x w(t,x)]^2 + \frac12 J(x) w(t,x)^2 \bigg] .
\end{equation}
Consequently, the Lagrangian in the continuum limit takes the form of 
\begin{equation}
    L = \int dx \; \frac12 \bigg[ (\partial_t w)^2 -K(\partial_x w)^2  - J   w^2 \bigg]. \label{eq:actionex}
\end{equation}

\begin{figure}[b] 
\centering
\begin{tikzpicture}[scale=0.7, every node/.style={scale=0.7}]

 \node[circle,draw,thick,minimum size=1.3cm,fill=gray!9] (H0) at ( 0,   0) {$h_{4N+3}$};
  \node[circle,draw,thick,minimum size=1.3cm,fill=gray!9] (H1) at ( 0,  -3) {$h_{7}$};
  \node[circle,draw,thick,minimum size=1.3cm,fill=gray!9] (H2) at ( 0,  -5) {$h_{3}$};
  \node[circle,draw,thick,minimum size=1.3cm,fill=gray!9] (H3) at ( 0,  -7) {$h_{-1}$};
  \node at (-3,-2.5) {\huge $\displaystyle \vdots$};
  \node at (3,-2.5) {\huge $\displaystyle \vdots$};
  \node at (0,-1.5) {\huge $\displaystyle \vdots$};
  \node[circle,draw,thick,minimum size=1.3cm,fill=RoyalBlue!6]            (X1) at (-3, -1) {$X_{4N+1}$};
  \node[circle,draw,thick,minimum size=1.3cm,fill=pink!18] (Y1) at ( 3, -1) {$Z_{4N+1}$};

  \node[circle,draw,thick,minimum size=1.3cm,fill=RoyalBlue!6]            (X2) at (-3, -4) {$X_{5}$};
  \node[circle,draw,thick,minimum size=1.3cm,fill=pink!18] (Y2) at ( 3, -4) {$Z_{5}$};

  \node[circle,draw,thick,minimum size=1.3cm,fill=RoyalBlue!6]            (X3) at (-3, -6) {$X_{1}$};
  \node[circle,draw,thick,minimum size=1.3cm,fill=pink!18] (Y3) at ( 3, -6) {$Z_{1}$};

  \draw[->,thick] (X1) -- (H0) node[midway, fill=white] {\small $W_{4N+2}^{(1)}$};  
  \draw[->,thick] (X2) -- (H1)  node[midway, fill=white] {\small $W_{6}^{(1)}$};
  \draw[->,thick] (X2) -- (H2) node[midway, fill=white] {\small $W_{4}^{(1)}$};
  \draw[->,thick] (X3) -- (H2) node[midway, fill=white] {\small $W_{2}^{(1)}$};  
  \draw[->,thick] (X3) -- (H3) node[midway, fill=white] {\small $W_{0}^{(1)}$};
  \draw[->, thick] (X1) -- (H1) node[midway, fill=white] {\small };
  \draw[<-,thick] (Y1) -- (H0) node[midway, fill=white] {\small $W_{4N+2}^{(2)}$};  
  \draw[<-,thick] (Y1) -- (H1) node[midway, fill=white] {\small };
  \draw[<-,thick] (Y2) -- (H1) node[midway, fill=white] {\small $W_{6}^{(2)}$};
  \draw[<-,thick] (Y2) -- (H2) node[midway, fill=white] {\small $W_{4}^{(2)}$};
  \draw[<-,thick] (Y3) -- (H2) node[midway, fill=white] {\small $W_{2}^{(2)}$};
  \draw[<-,thick] (Y3) -- (H3) node[midway, fill=white] {\small $W_{0}^{(2)}$};
  \node[rectangle, draw=white, fill=white, minimum size= 1cm ] at (1.5,-2) {\Large $ \cdots$};
  \node[rectangle, draw=white, fill=white, minimum size= 1cm ] at (-1.5,-2) {\Large $ \cdots$};

\end{tikzpicture}
\caption{Neural network architecture for the second example. Periodic boundary conditions are imposed.}
\label{fig:2nd example}
\end{figure}

As a second example, we examine a more realistic case that differs from the previous one in two respects. First, we insert an additional hidden layer between the input and output layers. Second, we adopt a quadratic activation function, $\sigma(x)=px^{2}+qx+r$. As before, we impose periodic boundary conditions and relabel the synapse indices, as illustrated in Fig.~\ref{fig:2nd example}.

Because the network depth in this example is small, we need not take a continuum limit in the depth direction, unlike in Eq.~\eqref{eq:clgen}. Instead, we associate each layer with a distinct field: $W_{1}$ and $W_{2}$ correspond to $w_{1}$ and $w_{2}$, respectively. When the depth becomes large, however, one can apply the continuum limit to the depth index. In that case, the collection $w_{1}(t,x),\,w_{2}(t,x),\,\ldots$ can be regarded as a single unified field $w(t,\mathbf{x})$, where $\mathbf{x}$ now includes the coordinate associated with depth.

The cost function for this neural network is
\begin{multline}
    C=\sum_{i,l} \Big[Y_{4i+1}-W_{4i}^{(2)}\sigma( W_{4i}^{(1)}X_{4i+1}+W_{4i-2}^{(2)}X_{4i-3}) \\
    -W_{4i+2}^{(2)}\sigma(W_{4i+2}^{(1)}X_{4i+1}+W_{4i+4}^{(1)} X_{4i+5} )  \Big]^2,
\end{multline}
and by following the procedure of Eq.~\eqref{eq:continuum}, we obtain
\begin{equation}\label{eq:second ex}
\begin{split}
    L =   &  \int   dx \    \Big[
    \frac{1}{2}(\partial_t w_1)^2+\frac{1}{2}(\partial_t w_2)^2 -\frac{1}{2} m^2 w_2^2 \\ & \hspace{5mm}  
    -J_1 - J_2 w_2 - J_{3} w_2 w_1 - K_{1} w_2 \partial_x w_1 \\& \hspace{5mm}-K_2w_2 \partial_x^2 w_2 
   - K_3 w_2 \partial_x^2 w_1   -K_4 \partial_xw_2 \partial_xw_1 \hspace{-5mm} \\& \hspace{5mm}  - K_5w_1\partial_xw_2-K_6\partial_x^2 w_2 - K_7 w_1\partial_x^2 w_2 +\cdots  \Big]. \hspace{-5mm}
    \end{split}
\end{equation}
Only terms up to quadratic order in $w_{1}$ and $w_{2}$, and up to second-order spatial derivatives, are shown. The coupling constants and external sources $m$, $K_{i}$, and $J_{i}$ are listed in Table~\ref{tab:coup}.

\vspace{2mm}
\textit{Discussion and Outlooks---} By establishing a concrete bridge between neural networks and field theory, our framework allows neural networks to be analyzed using field-theoretic language. For instance, in Eq.~\eqref{eq:second ex}, mass terms appear only for the synaptic field of the last layer $w_2$. This is because those parameters are the only ones that can couple to the constant term of the activation function.  This feature remains even when additional layers are added in the same way. By analyzing this mass term, the effect of the activation function's value at zero may be investigated. Likewise, further research could explore how specific features of neural networks shape the field-theoretic structure, or vice versa. Several open questions and directions for future work are listed below as examples of such possibilities.

(i) Network structures and properties: As noted above, understanding how the structural features of a neural network are encoded in its synaptic field theory is an intriguing open problem. Elements such as the training data set and the activation function can leave distinctive imprints on the resulting field theory. One could also study training protocols in which the data set or learning rate $\eta$ varies with time; the corresponding theory would then contain time‑dependent sources or exhibit expansion histories other than dS—e.g., radiation‑dominated evolution.
In parallel, developing a synaptic field theory that preserves locality will require a systematic indexing scheme applicable to more general architectures, including networks with complex connectivity, non‑trivial activations, and explicit bias terms. For example, if a network’s energy or cost function already encodes locality‑like structure \cite{hopfield1982neural, LITTLE1974101, Smolensky1986information, DynamicNeuronApproach}, that locality may carry over to the associated field theory.

(ii) Networks with conventional symmetries: Identifying architectures whose continuum limits respect familiar field‑theoretic symmetries, such as Poincaré invariance, would be illuminating. Our current constructions generically violate Poincaré invariance because these symmetries are not required in deep learning. Determining the conditions under which they emerge remains a challenging and intriguing problem.

(iii) Statistical‑physics perspective: Adding stochastic noise to the gradient‑descent dynamics turns the update rule into a Langevin equation, allowing a Gibbs‑measure description in which the cost function plays the role of a Hamiltonian. Although noise is absent from our present discussion, incorporating it should naturally link to earlier works \cite{ringel2025statistical} and provide a rich statistical field‑theoretic framework.

\begin{table}[t]
\centering
\renewcommand{\arraystretch}{1.5}
\begin{tabular}{l@{\hskip 20pt}l}
\hline
$m^2 = 8a^{-1}N_l r^2$ & \\
\hline
$J_1(x) = a^{-1} \sum_l (Y^{[l]})^2$ & $J_2(x) = 4ra^{-1} \sum_l Y^{[l]}$ \\
\multicolumn{2}{l}{$J_3(x) = 8qa^{-1} \sum_l ( X^{[l]} + 4a^2 \partial_x^2 X^{[l]} ) Y^{[l]}$} \\
\hline
$K_1(x) = 48aq \sum_l Y^{[l]} \partial_x X^{[l]}$ & $K_2 = 4a N_l r^2$ \\
$K_3(x) = 20qa \sum_l X^{[l]} Y^{[l]}$ & $K_4(x) = 16aq \sum_l X^{[l]} Y^{[l]}$ \\
$K_5(x) = 16aq \sum_l Y^{[l]} \partial_x X^{[l]}$ & $K_6(x) = 2ra \sum_l Y^{[l]}$ \\
$K_7(x) = 4qa \sum_l X^{[l]} Y^{[l]}$ & \\
\hline
\end{tabular}
\caption{Summary of couplings and external sources in Eq.~\eqref{eq:second ex}. Here, $a$ and $N_l$ denote the lattice spacing and the size of the training examples, respectively.}
\label{tab:coup}
\end{table}

\vspace{2mm}
\textit{Acknowledgements---} We thank KS for gently pointing out their earlier work, Ref.~\cite{Krippendorf:2022hzj}, which first discussed a duality connecting
neural network and cosmological dynamics. Regrettably, we were unaware of this reference when preparing the initial preprint of our manuscript. We thank J. Lee for his helpful comments. This work was supported in part by the National Research Foundation of Korea (Grant No. RS-2024-00352537).

\bibliography{ref.bib}

\end{document}